\documentclass{llncs}
\usepackage{makeidx}  
\usepackage{graphicx}
\begin{document}
%
%
%
\title{Towards a theory for Bio$-$Cyber Physical Systems Modelling}

\author{Didier Fass\inst{1} \and Franck Gechter\inst{2}}

\institute{ICN Business School, \\Mosel Loria UMR CNRS 7503, \\Universit\'{e} de Lorraine,\\ email: didier.fass@loria.fr
\and
IRTES-SET, EA 7274,\\ F-90010 Belfort Cedex,\\ email: franck.gechter@utbm.fr}


\maketitle


\begin{abstract}
Currently, Cyber–Physical Systems (CPS) represents a great challenge for automatic control and smart systems engineering on both theoretical and practical levels. Designing CPS requires approaches involving multi-disciplinary competences. However they are designed to be autonomous, the CPS present a part of uncertainty, which requires interaction with human for engineering, monitoring, controlling, performing operational maintenance, etc. This human-CPS interaction led naturally to the human in-the-loop (HITL) concept. Nevertheless, this HITL concept, which stems from a reductionist point of view, exhibits limitations due to the different natures of the systems involved. As opposed to this classical approach, we propose, in this paper, a model of Bio-CPS (i.e. systems based on an integration of computational elements within biological systems) grounded on theoretical biology, physics and computer sciences and based on the key concept of “human systems integration”. 
 
\end{abstract}

\section{Introduction}
Currently, researching theoretical principles of Cyber–Physical Systems (CPS) represents a great challenge for automatic control and smart systems engineering. Designing CPS requires a multidisciplinary approach involving mathematics, automatic control and applied mathematics and physics. Those disciplines are based on computation and regulation loop paradigms. However they are designed to be autonomous, one cannot neglect that the CPS present a part of uncertainty, which requires interaction with human for engineering, monitoring, controlling, performing operational maintenance, etc. This interaction led naturally to the human in-the-loop (HITL) concept, which is now widespread in literature. Nevertheless, this HITL concept has got several limitations due to the different natures of the systems involved and their different organizations. Indeed, the traditional system of systems engineering point of view stands from reductionism.  If this approach is suitable to computational sciences and physics, it is not well adapted to the human nature. This reductionist approach, i.e. mechanization and computerization of human being, reaches out its own limits and needs a scientific theoretical approach to face with future human machine challenges. 
In this paper, we propose a model of Bio-CPS grounded on theoretical biology, physics and computer sciences and based on the key concept of “human systems integration”. Bio-CPS are considered to be an integration of computational elements within biological systems. In a sense, Bio-CPS can be compared to the Cyber-Physical systems, in which the challenge is to make physical systems working along with computer systems. For decades, the Cybernetics has been a huge influence on human-machine systems concepts and development. It is based on information theory, automatic control theory, and algorithm theory. Cybernetics is about regulation and control of a mechanical system behavior. In this context, human-machine interactions are generally seen as an exchange of information between the operator and the controlled object. Hence, designing and developing human machines systems cannot be reduced to a simple information exchange. Using interactive and artificial technologies requires integrating artificial elements and structural design usually by artificial or artifactual functional interactions and its dynamics. Considering this, designing Bio-CPS is a real intellectual challenge since this can be considered as the “hyperlink”  between the biology and the cyber sciences. The development of new interactive systems for many applications such as airplane control, car smart driving assistance, bedside monitoring in intensive care, etc. and their growing implication in our daily life (smartphone, augmented and virtual reality, etc.), involve the necessity to develop theoretical basis for designing Bio-CPS.
This design task is hard to cope with. In addition to the classical issues encountered in Cyber-Physical Systems such as the relationship between a continuous time reference on the one side (physical part) and a dynamics based on sequences of state changes on the other (cyber part), Bio-CPS have add some other difficulties linked the differences in the nature of the interactions between system components with their scale relativity. Whereas the interactions between elements are, by nature, local and symmetrical in physical and, by extension, in cyber systems, they are non-symmetrical and non-local in biological systems. This difference in the nature of interaction is one of the main difficulties that have to be overcome. Another important issue is linked to the numerous hierarchical levels involved in biological systems from the cell level to the body level, the Physical or the cyber systems dealing at most with two levels. Finally, if the complexities of these two systems make the interoperability hard to design, the nature of these is also different in both cases. In biological systems the complexity is required in order to make the system more stable. By contrast, in cyber systems engineering, the complexity need to be avoided as far as it is possible because it can bring instabilities. However, cyber systems complexity, even not desired, stems from the numerous interactions between components. Consequently, the main concern in Bio-CPS design is how to make these two kinds of systems coupling together to perform a common task with a high level of confidence. 
The goal of this paper is to propose a general model aimed at helping the comprehension, the description and the design of Bio-CPS. The proposed design process of such systems integrates human factors and their fundamental ethological principles as the central element so as to obtain reliability and efficiency to the target common task to be performed. The paper will firstly present the main concepts involved and give some definitions linked to Bio-CPS. Then, after a detailed presentation of the related issues and of the possible applications, the paper will present a theoretical framework aimed at characterizing and designing such systems. 

\section{Concepts and definitions}

\subsection{Cyber-Physical Systems and the time representation issue}

Cyber-Physical Systems (CPS) are now widespread in literature and can be defined as follows:
\begin{definition}
CPS are systems with deep integrations of computational elements with physical processes (\cite{Lee10}).  
\end{definition}

To some points of view, it can be seen as an extension of the classical embedded systems which are mainly based on feedback loop concepts. Thus, one can consider the CPS as a mutual interaction between a computational processes, the time-line of which is based on discrete state changes, and a physical process based on continuous time evolution. The main issue of CPS is tied to the difficulty to manage this time-line duality. 

Two main CPS conception approaches can be found in literature whether one focuses on the cyber part or on the physical aspects. the "Cyberizing the Physical" approach \cite{Lee10} is based on the integration of computing element into physical process. This approach is generally linked to hybrid systems theories where the time is common to all the parts of the system. However, the model of time is still a problem because the time continuum linked to the physical system needs to be discretized to fit cyber requirements. If this discretization can be sufficient for physics inspired artificial intelligence systems for instance \cite{Gechter2011},  it is not the case for events based systems for which the use a superdense time model is required \cite{Manna93}. This time model is aimed to manage both continuous time-line and the causally related actions. 

By contrast, the "Physicalizing the cyber" approach \cite{Lee10} focuses on the integration of physical elements into computer science algorithms. This requires to re-think the abstractions used in classical approaches, which have now to fit physical part constraints and to reduce, as far as possible, the time variabilities and unpredictability due to the use of high level algorithms. Within these approaches the time issue is still the first order element. If the habit in computer science is the optimisation of the algorithms/computer couple so as to obtain the fastest possible behaviour, this point of view is no longer pertinent for CPS. Indeed, the efficiency of the computer side is not linked to the speed of its time response but to the adaptability of this response to the evolution of the physical part. Consequently, instead of being considered as a quality factor, the time becomes a semantic property common to both physical and cyber parts. Thus, the quality measurement of algorithms for CPS is related to adaptability, reliability, robustness, predictability, accuracy or repeatability. For these reasons, novel approaches such as parallel computing \cite{Jozwiak2014}, distributed computing \cite{Zhou2013}, multi-agent systems \cite{Dafflon2013}, etc. are well adapted to CPS issues.

\subsection{Complexity, emergence and complex systems}

A complex system can be defined as a set of a huge number of interacting entities, the global behaviour of which cannot be predicted by calculations or by an external observer. Generally, the evolution of the system is also unpredictable. Thus, a system is said as complex if the global obtained result (we'll see later in this paper what kind of result can be expected) can only be predicted by experiments and simulations even with a total knowledge of all its components and the rules that gather them. The existence of such systems challenges the reductionist approach \cite{Nagel61} which considers that the complex nature of systems can be reduced to a sum or a composition of fundamental principles. The complex system study is an activity widespread among many scientific fields.\\

Basically, there's a confusion between complicated system and complex system concepts. If one goes back to the etymological origin of these concepts, one will find the following definitions.

\begin{definition}
A system is said \textbf{complicated} if time and talent are required to understand it well. For instance, a clockwork mechanism is a complicated system. The complicated system organization is deterministic and computable.
\end{definition}

\begin{definition}
The term \textbf{complex} means that the system is made of many intrications which make impossible the study of a part of it separately while neglecting its other components. Even if some part of a complex system could be computable, it is mainly non-deterministic and unpredictable.
\end{definition} 
  
In \cite{Moncion2010}, a difference between these two concepts is made based on the dynamical nature of the relations between system components. Thus, even if both can be defined as a set of numerous interacting elements, the system components are considered to be fixed in time and space in a complicated system while they can vary dynamically into a complex system.

The complex nature of a system leads to two other concepts: the emergence and the self organizing ability. These concepts are closely tied and complementary. The self organizing ability is often linked to an increase in the order of the system or to a decrease of its entropy without any external control in computation sciences and automation. In \cite{DEWOLF2004}, the self-organization is defined as a dynamical and adaptive process allowing the system to obtain or to maintain an organizational structure without any external intervention. Therefore if the self organization concept is relevent for interactive systems then it remains inadequate for describing physical systems behaviour. 

If the self organization concept is clear and its definition is widely recognized with a certain scientific consensus, this is no truly the case for the emergence concept. Indeed, there exist many different definitions for emergence. In the common mind, the emergence concept is linked to an existing external observer who is able to determine and to analyse the phenomena produced by a process. During the XIX$^th$ century, has been introduced in biology and in philosophy as opposed to reductionism. Then, more recently, this concept lead to a great deal of research work in several scientific domains such as in computer science, \cite{DESSALLES2008}, \cite{Bonabeau95a}, \cite{Bonabeau95b}, \cite{Yamins05a}, \cite{Yamins05a}, in complex systems study \cite{Moncion2010}, in sociology,...

By the same as for the concept of complex system which cannot be studied into a reductionist frame, le concept of emergence share the same kind of problems. In \cite{Memmi96} for instance, emergence is reduced to a simple problem of description and explanation. In this case, it is not a system property but a property of the point of view one may have on the system. In \cite{DEWOLF2004}, a system is defined to present emerging properties when phenomena appear dynamically at a macroscopic level as a result of interactions between system components which occur at a microscopic level. In \cite{BEDAU97} and \cite{BEDAU2002}, emergence concept is split into three categories:

\begin{itemize}
\item The \textbf{nominal emergence}: it is linked to the presence of macroscopic properties which can not be defined to be microscopic. 
\item The \textbf{weak emergence}: it can be considered as a sub part of the nominal emergence where the appearance of the phenomena cannot be explained easily. As said in \cite{BEDAU97}, this type of emergence requires the use of simulations and experiments.
\item The \textbf{strong emergence}: as opposed to the nominal emergence, the strong emergence consider that the observed phenomena on a macroscopic point of view have side effects on both macroscopic and microscopic levels.
\end{itemize}

Considering these definitions, it is now clear that the self-organizing ability is tied to the organisational structure of the system whereas the emergence of properties or functions is tied to the dynamical aspect of the system. \\

For us self-organization and emergence are system properties resulting from the interactions between the cyber and physical parts of the system.

\subsection{Interactions: the central elements in CPS}

Interactions are the central elements for modeling CPS . Their nature and their conception trigger the way the coupled global system evolves and how it is able to reach its goal(s).


The general definition of an interaction is the following:

\begin{definition}
An interaction is a dynamical relationship between two or more entities based on a set of reciprocal actions.
\end{definition}

Interactions can have different forms. Interactions can be direct,i.e. one entity affects directly another one by an action (collision between two physical elements for instance). This kind of interactions is generally encountered in physical systems. By contrast, interactions can be also indirect, i.e. the action of one entity is propagated to another one by using an intermediate element. This element can be a part of the entities common environment or another part of the system. These interactions are mainly met in social biological systems (ant colonies \cite{Parunak97}, social spiders, etc.) and can lead to stigmergy \cite{Theraulaz1999} (i.e. a form of self-organization/coordination mechanism induced by mutual actions performed on a shared element).
Besides this direct/indirect properties, some other properties are crucial for Bio-CPS coupling. Among them, the local and the symmetrical characters of interactions are particularly pertinent.

\begin{definition}
An interaction is said to be local when the mutual actions between elements are performed within a short distance. By contrast, an interaction is said non-local when it is performed between elements separated in space with no noticeable intermediate agency or mechanism.
\end{definition}

Among local interactions, we can cite as example the local application of forces in Physics such as the tyre/road interaction while studying vehicle behaviours. Non-local interaction can also be found in Physics (the gravitation law or the Aharonov–Bohm effect for instance \cite{Figielski1998} are common examples) but are mainly present in biological systems ruling the exchanges between organs in human bodies for instance \cite{Chauvet1993c}. One other important character of interactions is the symmetrical aspect.

\begin{definition}
An interaction is said symmetrical when there is a reciprocity between the elements involved.
\end{definition}

In physical world, interactions are generally symmetrical (the electrical interaction between particles for instance) whereas in biological system the symmetry is not the common rule \cite{Chauvet1993b}. However, some research works are starting to use virtual Physics inspired interaction  where the symmetric character has been removed. For instance, \cite{Contet2011} and \cite{Ali2015} are proposing models of non-symmetric physical interactions between vehicles in platoon in order to increase the stability of the system as compared to symmetric spring damper functions.


\section{Bio-CPS issues and applications}

Today human artefact systems are a design and engineering challenge. From human machine system to sociotechnical, systems automation and interaction ground many technical development. Traditional approaches rely on analytic and a reductionist method. They propose an abstraction of the human based on a mechanical or computational model. The limits of that epistemological metaphor leads to the necessity of a strong paradigm shift.

When one wants to make human and CPS working together so as to accomplish a task, the question of the nature of the relationship has to be addressed. Does it corresponds to an \textit{interaction} or to a \textit{coupling} between two systems? The problem is the fact that the two systems are different by nature. Even if they are embedded into a common physical space (i.e. a place where the physical principles are respected); the are different in their structural and dynamic organization. 

The challenge is then to be able first to find out an isomorphic framework for describing the two systems and their integrative coupling with a shared reference frame, and secondly to validate fundamental human cyber-physical systems integration principles.

So the Bio-CPS modeling needs bio-compatible and bio-integrable CPS design and engineering scientific grounding.  

\subsection{Human and machine : different natures, one isomorphic framework }

\subsubsection{Two different natures}

The human nature is biological (and anthropological, grounded on the former). The machine nature is artificial and cyber-physical. These evidences constrain to define a new theoretical and experimental framework for describing, designing and synthesizing Bio-CPS.

Thus one of the main issues is to think Bio-CPS as a global system of one higher or of one different level of organization of the human body. CPS may be modeled as an anatomical and physiological extension of the biological system and then as an enhancement of its own and social domain of activity and life.

\subsubsection{Interaction}
One of the main organizational difference from biological systems and CPS is the interaction nature.

As previously said, the interactions in both cyber and physical worlds are all symmetric and mostly local. Some interactions can be non-local but they are focused on physical side and are not met into cyber-physical set of interactions. The reason of this is embedded in the regulation loop concept which is the main inspiration source for CPS.

By contrast, biological interactions are functional interactions. This theoretical concept of functional interaction (i.e. something emitted by an entity has got an action upon on another entity somewhere) introduced by Chauvet \cite{Chauvet1993a}, \cite{Chauvet1993b}, \cite{Chauvet1993c} explains the specific nature of the human interactions. Their three main properties are : non-symmetric, non local and non-instantaneous.

At this point, one can see the difficulty in making biological systems and cyber-physical systems interacting together so as to perform a common task or to reach one common goal as a global entity. Modeling and engineering a Bio-Cyber Physical system relies on making three complex systems, with different interaction natures and with different time representations, working together.

Until now, the willing of interactions between human and CPS has led naturally to the human in-the-loop (HITL) concept, which is now widespread in literature. Nevertheless, this HITL concept has got several limitations due to the different natures of the systems involved and their different organizations. Indeed, the HITL concept stems from a traditional system of systems engineering point of view linked to reductionism concepts. In that epistemological context modeling or designing human machine interaction is representing interaction as closed symmetrical loop and behavioural process.\\

After the previous considerations on the nature of human and CPS interactions, it is now clear that the HITL concept has to be override by the introduction of new theoretical foundations. Table \ref{table1} presents one classification of biological, cyber and physical systems considering the nature of the interactions and their time representations.

\begin{table}
\begin{center}

\begin{tabular}{|c|c|c|c|}

\hline 
• & Biological & Cyber & Physical \\ 
\hline 
Symmetricity  & Never & Both symmetric  & Always \\ 
of interactions&   & and  non-symmetric &\\
\hline 
Locality & Mainly non-local & Mainly local & Both local  \\ 
of interactions & &&and non-local\\
\hline 
Time representation & Continuous (Functional level) & Discrete  & Continuous \\ 
& Discrete (Structural level)&or Event based&\\
\hline 
\end{tabular} 
\caption{Classification of systems considering the nature of the interactions and the time representation}
\label{table1}
\end{center}
\end{table}

\subsubsection{One isomorphic framework}
First to revise the HITL concept and to shift to our bio-integrative paradigm, we have proposed \cite{Fass2012} an isomorphic framework for modeling natural or artificial systems. This conceptual framework describes three categories of required main system dimension: structural elements, shapes or forms and dynamics. Taking into account two by two this main classes of system variables, one can describe three specification plan: architecture (structural elements to shape or form specify geometrical structure or system architecture), behaviour (shape or form to dynamics specify analytical functions or functionally analysed) and evolution (structural elements to dynamics specify three main types of functional interactions: physical $\Phi$, logical $\Lambda$, biological $\Psi$).

\begin{center}
\begin{figure}
\includegraphics[scale=0.3]{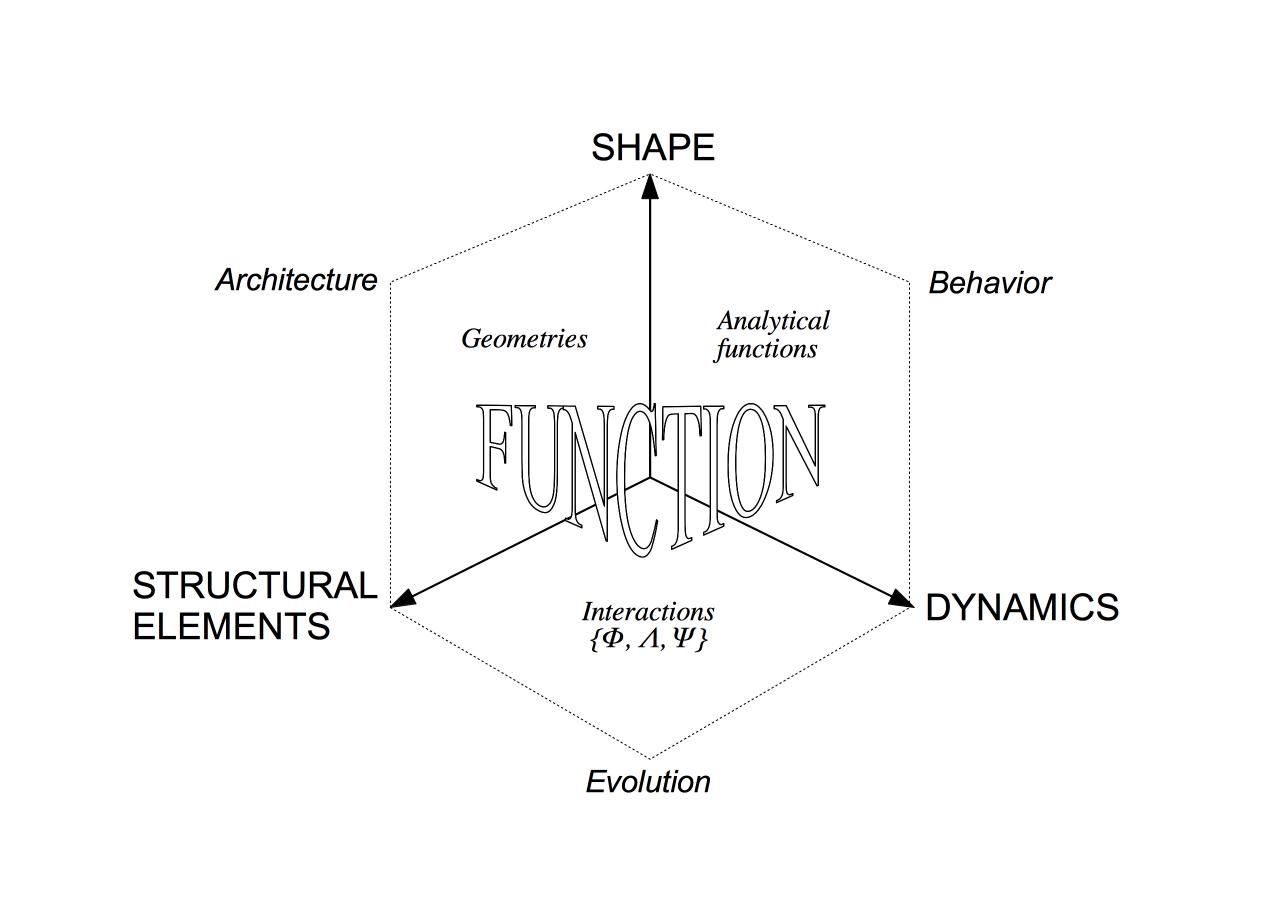}
\caption{The Bio-CPS isomorphic framework}
\end{figure}
\end{center}

If one one assume that function does not exist by itself but is the emerging result of integrative organization, this framework grounds our bio-integrative model based Bio-CPS engineering.





\section{Conclusion}

After explaining the main issues and the main differences in the natures of biological, physical and cybernetics systems, this paper presents a Bio-CPS framework aimed at helping in conceptualizing, designing and engineering Bio-CPS. 

The proposed approach consists in thinking up such systems by considering the CPS system as an extension and/or an enhancement of the human biological nature which requires bio-compatible and bio-integrable scientific and design principles. This approach is a paradigm shift as compared to the classical approaches which reduce the human to only mechanical, logical or computational properties.\\
As opposed to CPS, where the interactions are only logical and physical, the interactions in Bio-CPS can be considered as multi-modal couplings which involves physical, logical and biological modalities and specific combinations of these.\\
So as to ensure a complete and coherent coupling between systems in Bio-CPS, one must design and engineer interactions which deal with biological, physical and logical elements. The coupling being made at interaction level, the bio-compatibility must be ensured by the choices made for the structural elements, the dynamics and the form or shape and thus made for the global design of the Bio-CPS.\\
The approach, proposed in this paper, is epistemologically different from traditional CPS modelling method based only on a functional and behavioural analysis and metaphorical reductionism. Our Bio-CPS modelling is ground on both a structural analytical description and a functional integrative synthesis based on bio-compatibility and bio-integration needs and theoretical principles. In this three-dimensional integrative isomorphic framework - 3D-space of requirements and specifications, the interaction sub-space takes into account both statical and dynamical evolution of integrated Bio-CPS structural systems. The desire and emerging global function results from this integrative structural and dynamics organization. 
According this modelling approach human is not reduced to a logical or a logical factor or sub-system. CPS to be couple with human must fulfil biological requirements of human nature, its domain of life and activity.  

Thus we can plan to model and construct with correctness bio-compatible and bio-integrative CPS. This conception and design of Bio-CPS will ensure not only the coupling interactions but the reliability of the overall integrated function of the artificially extended human body and its domain of life and activity in the time. This is an issue for the design of safety critical systems next-gen CPS for extending and enhancing human capability in health, aerospace, transport and defense.

\section*{Acknowledgements}



%

\end{document}